\documentclass[aps,prl,twocolumn,showpacs,groupedaddress]{revtex4}
\usepackage{graphicx}

\begin{document}


\title{Noise limits in matter-wave interferometry using degenerate quantum
gases.}
\author{Chris P. Search and Pierre Meystre}
\affiliation{Optical Sciences Center, The
University of Arizona, Tucson, AZ 85721}

\date{\today}

\begin{abstract}
We analyze the phase resolution limit of a Mach-Zehnder atom
interferometer whose input consists of degenerate quantum gases of
either bosons or fermions. For degenerate gases, the number of
atoms within one de Broglie wavelength is larger than unity, so
that atom-atom interactions and quantum statistics are no longer
negligible. We show that for equal atom numbers, the phase
resolution achievable with fermions is noticeably better than for
interacting bosons.
\end{abstract}

\maketitle

Atom interferometers \cite{berman} offer a potential sensitivity
that exceeds that of their optical counterparts by as much as
$Mc^2/\hbar\omega\sim 10^{10}$ \cite{clauser,scullydowling}.
Besides being useful for fundamental tests of quantum mechanics
and precision measurements of fundamental constants, this property
offers great promise for inertial navigation as well as for
geophysical applications, tests of general relativity, etc.
\cite{rotation1,gradiometer} In particular, recent years have witnessed
significant experimental progress towards the development of
atomic rotation sensors \cite{rotation1, rotation2, rotation3}
with sensitivities now surpassing those of optical gyroscopes
\cite{rotation1}. So far, experiments with atom interferometers in
the spatial domain have always been carried out in the
nondegenerate regime where quantum statistics and many-body
interactions are negligible and the interference pattern is the
sum of single atom interference patterns. This is true even in
those situations where Bose-Einstein condensates (BEC's) have been
used as sources: in such cases, the interference experiments were
performed after the atoms have been released from their trapping
potentials and allowed to undergo a ballistic expansion in order
to eliminate any mean-field effects, see for example
\cite{torii,gupta}. Likewise, the theoretical analysis of the
sensitivity of atom interferometers has also been confined to this
regime. Refs. \cite{scullydowling,dowling,yurke} analyzed the
quantum noise limit of a matter-wave Mach-Zehnder interferometer
and showed that for uncorrelated inputs the phase resolution is
$1/\sqrt{N}$, for bosons as well as for fermions. Here, $N$ is the
total number of particles entering the interferometer. The phase
resolution limit results from the intrinsic quantum fluctuations
associated with the sorting of the particles by the beam splitters
in the interferometer. The reason that bosons and fermions yield
the same sensitivity in these treatments is a consequence of the
assumption that the fermions are quasi-monochromatic, i.e. the
fermions all occupy the same state. This can only be justified
when the atomic wave functions do not overlap
\cite{scullydowling}.

Very much like the invention of the laser revolutionized optics
and interferometry, it would seem desirable to extend the
operation of matter-wave interferometer to truly
quantum-degenerate atom sources. It is therefore important to
extend these theoretical results to the case of highly degenerate
beams of bosons or fermions for which $\rho\lambda_{d}\gtrsim 1$,
where $\rho$ is the atomic density and $\lambda_{d}$ is the de
Broglie wavelength. In this limit, the effects of quantum
statistics and of atom-atom interactions are no longer negligible,
and it is expected that the sensitivity of the interferometer will
decrease. For bosonic atoms, the interference pattern and phase
sensitivity is degraded because of the atom-atom interactions,
which give rise to a nonlinear phase shift as the atoms propagate
along the arms of the interferometer. In contrast, for
spin-polarized fermions --- where $s$-wave scattering is absent
for $T\simeq 0$
--- the phase sensitivity is degraded because the Pauli exclusion
principle forces each of the atoms in the interferometer to occupy
a different momentum state. Our central result is that contrary to
what one might expect, the sensitivity of a Mach-Zehnder
interferometer using a degenerate beam of fermions can be
noticeably better than what can be achieved using a BEC.

Figure 1 shows the atomic Mach-Zehnder interferometer that we
consider, with input ports $A$ and $B$. For convenience we assume
that it can be treated as an effective one-dimensional system in
which the propagation of the atoms transverse to the
interferometer arms can be ignored. That is, we assume that the
atoms are tightly confined in the directions transverse to the
interferometer paths, with the same wave function $\phi_0({\bf
r}_{\perp})$. This holds when $\mu < \hbar\omega_{\perp}$ where
$\mu$ is the chemical potential for the bosons or fermions and
$\hbar\omega_{\perp}$ is the spacing between the energy levels of
the transverse wave functions, assumed to be harmonic. We
introduce the annihilation (creation) operators $\hat{a_k}$
$(\hat{a_k}^\dagger)$ and $\hat{b_k}$ $(\hat{b_k}^\dagger)$ for
atoms entering the input ports $A$ and $B$, respectively, with
momentum $\hbar k$ directed along either the upper or lower
interferometer path. They obey the commutation relations
$\hat{a}_k\hat{a}_{k'}^\dagger \pm
\hat{a}_{k'}^\dagger\hat{a}_k=\delta_{k,k'}$ and
$\hat{b}_k\hat{b}_{k'}^\dagger \pm
\hat{b}_{k'}^\dagger\hat{b}_k=\delta_{k,k'}$ where the minus sign
is for bosons and the plus sign is for fermions.

\begin{figure}
\includegraphics*[bb=0 0 425 625,width=5cm,height=8cm,angle=-90]{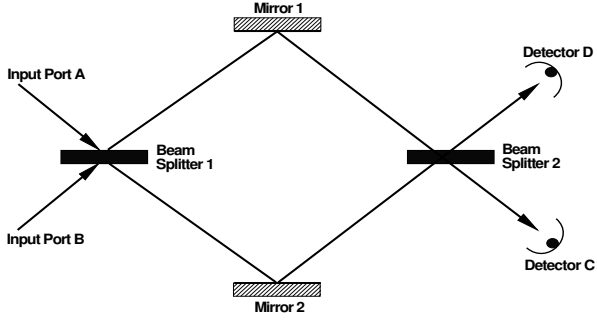}
\caption{Mach-Zehnder interferometer. $\ell_u$ and $\ell_l$ are the path 
lengths for the upper and lower arms, respectively. }
\label{fig1}
\end{figure}

The beam splitters 1 and 2 are identical 50-50 beam splitters,
which in the case of the first beam splitter has the action
\[ \left(\begin{array}{c} \hat{a}_{u,k} \\ \hat{a}_{l,k}
\end{array} \right)=\frac{1}{\sqrt 2} \pmatrix{&i &-1 \cr &-1 &i}
\left( \begin{array}{c} \hat{a}_k \\ \hat{b}_k
\end{array} \right)
\]
where $\hat{a}_{u,k}(\hat{a}_{l,k})$ are the operators for the
upper (lower) paths of the interferometer. The mirrors $M_1$ and
$M_2$ each result in a $\pi/2$ phase shift. The Hamiltonian
governing the evolution of the Heisenberg operators in the two
branches of the interferometer is
\[
H=\sum_{s=l,u}
\sum_{k}\hbar
\left(\omega_{k}+\omega_{\perp}+\frac{1}{2}\chi(\hat{a}^{\dagger}_{s,k}\hat{a}_{s,k}
-1) \right) \hat{a}^{\dagger}_{s,k}\hat{a}_{s,k}
\]
where $\omega_k=\hbar k^2/2M$ and $M$ is the atomic mass. We
recall that $\chi \equiv 0$ for spin-polarized fermions at
$T\simeq 0$, while for bosons $\chi=(4\pi\hbar a/M\ell)\int d{\bf
r}_{\perp} |\phi_0({\bf r}_{\perp})|^4$ , $a$ being the $s$-wave
scattering length and $\ell\approx\ell_{l,u}$. The Hamiltonian
ignores momentum changing collisions, which for an initial state
consisting of a weakly interacting BEC amounts to neglecting the
quasiparticle fluctuations about the macroscopically occupied
condensate mode. This is equivalent to a single-mode approximation
for the condensate.

After the beams are recombined at the second beam splitter, the
atoms are counted at detectors located at the output ports $C$ and
$D$. The relevant observable is the difference in the number of
counts,
\begin{equation}
\hat{{\cal
N}}=\hat{N}_D-\hat{N}_C=\sum_{k}\left(\hat{d}^{\dagger}_k\hat{d}_k-
\hat{c}^{\dagger}_k\hat{c}_k \right)
\end{equation}
where $\hat{d}_k$ and $\hat{c}_k$ are the annihilation operators for the output
ports $D$ and $C$, respectively.

In what follows, we consider only uncorrelated inputs where the
initial quantum state can be written as the factorized product of
Fock states $|\Psi_A\rangle|\Psi_B\rangle$. In this case, we recall that for
noninteracting atoms occupying the same state, the expectation
value of $\hat{{\cal N}}$ is $\langle{ \hat {\cal N}
}\rangle=(N_A-N_B)\cos\theta$ where $\theta$ is the phase
difference between the two interferometer paths and $N_A$ and
$N_B$ are the number of atoms entering via the two input ports.
Hence, ${\hat {\cal N}}$ is a direct measure of $\theta$, with
fluctuations
\begin{equation}
\Delta\theta=\frac{\Delta {\cal N}}{|\partial\langle\hat{{\cal N}
}\rangle /\partial\theta |}
\end{equation}
and $\Delta {\cal N}\equiv [\langle\hat{{\cal N}
}^2\rangle-\langle\hat{{\cal N}}\rangle ^2]^{1/2}$
\cite{scullydowling,dowling,yurke,yurke2}. We now compare $\Delta
\theta$ for the cases of a single mode BEC and of a spin-polarized
Fermi gas.

{\em Bose-Einstein Condensate:} Here we consider $N_A$ bosons
entering port $A$ and $N_B$ entering port $B$ in a time $\tau$,
each atom having velocity $v_0=\hbar k_0/m$. As such, $\langle
\hat{{\cal N}} \rangle$ represents the number of atoms counted in
the interval $\tau$. The quantum state describing this incident
beam is
$$|\Psi(0)\rangle=\left(\hat{a}^{\dagger}_{k_0}\right)^{N_A}
\left(\hat{b}^{\dagger}_{k_0}\right)^{N_B}|0\rangle
/\sqrt{N_A!N_B!}.$$
We evaluate the moments of $\hat{{\cal N}}$ by using
an angular momentum representation for the bosonic operators
\cite{yurke2,sakurai},
$2\hat{J}_z=\hat{a}^{\dagger}_{k_0}\hat{a}_{k_0}-\hat{b}^{\dagger}_{k_0}\hat{b
}_{k_0}$,
$2\hat{J}_x=\hat{a}^{\dagger}_{k_0}\hat{b}_{k_0}+\hat{b}^{\dagger}_{k_0}\hat{a
}_{k_0}$,
$2i\hat{J}_y=\hat{a}^{\dagger}_{k_0}\hat{b}_{k_0}-\hat{b}^{\dagger}_{k_0}\hat{a
}_{k_0}$, and $\hat{J}^2=\hat{N}/2(\hat{N}/2+1)$ where
$\hat{N}=\hat{a}_{k_0}^{\dagger}\hat{a}_{k_0}+\hat{b}_{k_0}^{\dagger}\hat{b}_{k_0}$.
In this representation, the incident beam state is
$|\Psi(0)\rangle=|j,m\rangle$, where
$$j=(N_A+N_B)/2, \,\,\,\,\,
m=(N_A-N_B)/2.$$ After some algebra one finds
\begin{widetext}
\begin{eqnarray}
\langle{\hat {\cal N}}\rangle&=&2m d^{(j)}_{m,m}(\beta)\cos\theta
+ \left( \sqrt{(j-m)(j+m+1)}d^{(j)}_{m+1,m}(\beta) +
\sqrt{(j+m)(j-m+1)}d^{(j)}_{m-1,m}(\beta)\right) \sin\theta, \label{MBEC} \\
\langle{\hat {\cal N}}^2\rangle&=&j(j+1)+m^2+
f_{j,m}(2\beta)\cos(2\theta+\beta)+g_{j,m}(2\beta)\sin(2\theta+\beta),
\label{MBEC2}
\end{eqnarray}
where
\begin{eqnarray}
f_{j,m}(2\beta)&=&(3m^2-j(j+1))d^{(j)}_{m,m}(2\beta)-
\frac{1}{2}\sqrt{(j-m+2)(j+m-1)(j-m+1)(j+m)}d^{(j)}_{m-2,m}(2\beta) \nonumber \\
&-&\frac{1}{2}\sqrt{(j+m+2)(j-m-1)(j+m+1)(j-m)}d^{(j)}_{m+2,m}(2\beta) ,\\
g_{j,m}(2\beta)&=&(2m-1)\sqrt{(j-m+1)(j+m)}d^{(j)}_{m-1,m}(2\beta)+(2m+1)\sqrt{(j+m+1)(j-m
)}d^{(j)}_{m+1,m}(2\beta),
\end{eqnarray}
\end{widetext}
and $d^{(j)}_{m',m}(\beta)=\langle
j,m'|\exp(-i\beta\hat{J}_y)|j,m\rangle$ are the matrix elements of
a rotation about the $y$-axis by an angle $\beta=\chi (t_l+t_u)$ .
In these expression, $\theta$ is again the phase difference
between the two interferometer paths,
\begin{equation}
\theta=\left(\omega_{k_0}+\omega_{\perp}+\chi(N_A+N_B-1)/2\right
) (t_u-t_l) + \beta/2, \label{phase}
\end{equation}
where $t_l=2\ell_l/v_0$ ( $t_u=2\ell_u/v_0$) is the transit time
for a point of constant atomic phase through the lower (upper) arm
of the interferometer. Note that when $\chi\equiv 0$, 
$d^{(j)}_{m',m}(0)=\delta_{m',m}$ and our results agree with prior calculations\cite{yurke2}. 

The nonlinearity has a two-fold effect on the interference 
pattern. This is seen by noting that the phase acquired by an atom 
propagating along the upper or lower path is 
$\hat{\phi}_s=(\omega_{k_0}+\omega_\perp+\chi\hat{a}^{\dagger}_{s,k_0}\hat{a}_{s,k_0})t_s$, 
and includes a self-phase modulation due 
to the other atoms traveling along the same path. However, the wave functions of the atoms 
entering the interferometer are coherently split between the two paths so that the 
effect of the nonlinearity is no longer a simple self-phase modulation when 
expressed in terms of the incident quantum fields,
\[
\hat{\phi}_u-\hat{\phi}_l=(\omega_{k_0}+\omega_{\perp}+\chi\hat{N}/2)(t_u
-t_l)-\chi (t_u+t_l)\hat{J}_y.
\] 
The phase difference includes a self-phase modulation 
proportional to half the total number, a simple consequence of the fact that 
the atoms are equally split between the two paths. However $\hat{\phi}_s$ also includes 
a term proportional to $\hat{J}_y$. This results from the fact that the wave 
functions of the atoms entering the interferometer are {\em coherently} split between the two 
paths at the first beam splitter. Hence, the $\hat{J}_y$ contribution is a result of the 
indistinguishability of the bosons that enter via the two input ports.

We remark that $\theta=\langle \hat{\phi}_u-\hat{\phi}_l \rangle 
+\beta/2-\chi(t_u-t_l)/2$ where the last two terms, $\beta/2-\chi(t_u-t_l)/2$, can be interpreted as a 
"vacuum effect", resulting from the non-commutativity of $\hat{a}_{s,k_0}$ and 
$\hat{a}^{\dagger}_{s,k_0}$. Its main impact on $\theta$ is to replace 
$N\equiv N_A+N_B$ with $N-1$, showing explicitly that the nonlinear phase shift 
vanishes if only one atom is present. The $\beta/2$ simply shifts the fringe 
pattern and is included in the definition of $\theta$ for convenience. 

It is clear from Eqs. (\ref{MBEC}) and (\ref{MBEC2}) and the definition of
$d^{(j)}_{m',m}(\beta)$ that $\beta j$ is the critical parameter
that determines the effect of the atom-atom interactions on
$\Delta\theta$. Relevant values of $\beta$ can be estimated by
assuming a typical chemical potential of $\mu/\hbar \sim
10^3-10^4s^{-1}$ and that $\phi_0({\bf r}_{\perp})$ is the
harmonic oscillator ground state for
$\omega_{\perp}\gtrsim\mu/\hbar$. Taking the perimeter of the
interferometer to be $\ell_l+\ell_u \sim$ 1 cm and typical recoil
velocities, $v_0\sim$ 1-10 cm/s, gives $\beta \gtrsim 10^{-3}$
for $^{87}$Rb or $^{23}$Na, so that $\beta j \gtrsim 1$ typically.

The results of a numerical evaluation of Eqs. (\ref{MBEC}) and
(\ref{MBEC2}) are shown in Figs. 2 and 3. We find that
$\Delta\theta$ is minimum at $\theta=(2n+1) \pi/2$ and exhibits
sharp resonances at $\theta=n\pi$ where $n$ is an integer. These resonances 
disappear in the limit $\beta\rightarrow 0$. These features agree qualitatively 
with a perturbation expansion for $\Delta\theta$ to lowest order in $\beta j$,
$\Delta\theta=\sqrt{\left[j(j+1)-m^2\right]/2m^2}\left(1+(\beta 
j)^2(a(m,j)+b(m,j)\cot^2\theta)\right)$
where $a(m,j)$ and $b(m,j)$ are positive definite and of order unity. 
Fig. 2 shows that for fixed $N_A$ the normalized phase fluctuations, 
$\Delta\theta(\beta)/\Delta\theta(\beta=0)$, increase with increasing
$N_B$ and for fixed $N$, the nonlinearity also results in
an increase in the phase fluctuations with increasing $N_B$. Figure 3 shows that for
$\beta\lesssim 10^{-3}$, $\Delta\theta$ continues to scale as
$1/\sqrt{N_A}$ when $N_B=0$ while for $\beta>10^{-3}$,
$\Delta\theta$ can actually show an increase with increasing
$N_A$.

\begin{figure}
\includegraphics*[width=8cm,height=5cm]{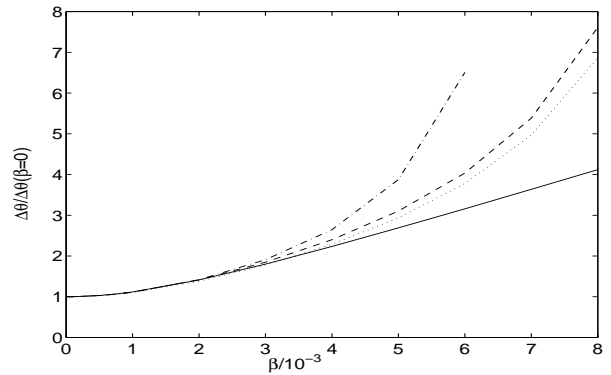}
\caption{$\Delta\theta$ at $\theta=\pi/2$ as a function of $\beta$ for $N_A=1000$ and
$N_B=0$ (solid line), $N_A=1000$ and $N_B=50$ (dashed line), $N_A=1000$ and
$N_B=100$ (dash dot line), and $N_A=950$ and $N_B=50$ (dotted line).}
\label{fig2}
\end{figure}

\begin{figure}
\includegraphics*[width=8cm,height=5cm]{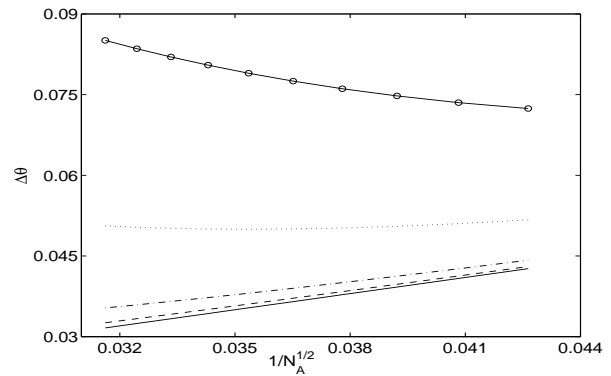}
\caption{$\Delta\theta$ at $\theta=\pi/2$ as a function of
$1/\sqrt{N_A}$ for $N_B=0$ and $\beta=0$ (solid line), $\beta=0.0005$ (dashed
line), $\beta=0.001$ (dashed dot line), $\beta=0.0025$ (dotted line),
$\beta=0.005$ (circles).}
\label{fig3}
\end{figure}

{\em Fermi Gas:} As with the bosons, we consider uniform beams of
$N_A$ and $N_B$ fermions entering the interferometer via input
ports  $A$ and $B$ in a time $\tau$ at average velocities $v_0$.
Their densities are $n_A=N_A/v_0\tau$ and $n_B=N_B/v_0\tau$, and the 
quantum state for the incident fermions is then given by
\[
|\Psi(0)\rangle=\prod_{|k-k_0|\leq k_{FA}}\hat{a}^{\dagger}_k\prod_{|k-k_0|\leq
k_{FB}}\hat{b}^{\dagger}_k|0\rangle
\]
where $k_{FA}=\pi n_A$ and $k_{FB}=\pi n_B$ are the Fermi wave
numbers for the two beams. We note from the definitions of
$k_{FA}$ and $k_{FB}$ that there exists a new fundamental length
scale, $L_F=v_0\tau$ that is important for the description of the
fermions but was unimportant for the bosons. It represents the
length of the two incident beams and can be interpreted as the
quantization volume for the fermions.

$\langle \hat{{\cal N}}\rangle$ and $\Delta {\cal N}^2$ are now found to
be
\begin{eqnarray}
\langle\hat{{\cal N}}\rangle&=&\cos\theta\left[
\bar{n}_AS(\bar{n}_A\theta/2)-\bar{n}_BS(\bar{n}_B\theta/2)\right]/\alpha
\label{MFERMI} \nonumber \\
\Delta {\cal N}^2&=&\left( \bar{n}_A-\bar{n}_B-\cos 2\theta
\left[\bar{n}_AS(\bar{n}_A\theta)-\bar{n}_BS(\bar{n}_B\theta)
\right] \right)/2\alpha \nonumber
\end{eqnarray}
where $S(x)=\sin(x)/x$, $\alpha=(2\pi/k_0)/L_{F}$,
$\theta=k_0(\ell_u-\ell_l)$ is the phase difference obtained for a
single atom with momentum $\hbar k_0$, and we have taken $N_A>N_B$
without loss of generality. We have also introduced the
dimensionless densities $\bar{n}_A=2\pi n_A/k_0=2k_{FA}/k_0$ and
$\bar{n}_B=2\pi n_B/k_0=2k_{FB}/k_0$. They represent the number of
atoms within the de Broglie wavelength $\lambda_0=(2\pi)/k_0$, so
that quantum degeneracy corresponds to $\bar{n}\gtrsim 1$. Note
that $\bar{n}$ is also a measure of the degree of monochromaticity
of the incident Fermi beam. Earlier treatments of fermion
interferometers dealt with quasi-monochromatic beams that
correspond to $\bar{n}\ll 1$.

We first restrict ourselves to the case $\bar{n}_B=0$. The phase
uncertainty is then
\begin{eqnarray}
&&\Delta\theta =\frac{1}{\sqrt{2N_A}} \nonumber \\
& \times & \frac{\sqrt{\theta^2(1-\cos 2\theta S(\bar{n}_A\theta))}}
{\left|\cos\theta\cos(\theta\bar{n}_A/2)-S(\bar{n}_A\theta/2)[\theta\sin\theta+
\cos\theta]\right|}, \label{DTF}
\end{eqnarray}
which shows that $\Delta\theta\sim 1/\sqrt{N_A}$. However, Eq.
(\ref{DTF}) also indicates that $\Delta\theta$ is an increasing
function of $\theta$ and hence of the path length difference. This
is easily understood by noting that the individual phases
contributing to the total interference pattern are uniformly
distributed over the interval
$[\theta-\delta\theta,\theta+\delta\theta]$, where
$\delta\theta=k_{FA}|\ell_u-\ell_l|=\bar{n}_A|\theta|/2$.
Consequently, for degenerate fermion beams with $\bar{n}_A\gtrsim
1$, the interference pattern is undetectable for large path
differences, $|\theta|\gtrsim \pi$. This will continue to hold when $\bar{n}_B\neq 
0$.

Figure 4 shows the minimum value of $\Delta\theta$ for fixed
$\bar{n}_A$ and various $\bar{n}_B$ plotted as a function of
$\sqrt{\alpha}\sim 1/\sqrt{N}$. It is clear that the phase
fluctuations continue to scale as $1/\sqrt{N}$. Moreover, a
comparison with Fig. 2 shows that increasing the number of atoms
entering via input port B does not result in as dramatic an
increase in $\Delta\theta$ as with bosons. Whereas for bosons
having only $\sim 10\%$ of the atoms enter via port B could result
in an increase in $\Delta\theta$ by a factor of 5 or more for
moderate values of $\beta$, having more than a third of the atoms
enter via B now results in an increase in $\Delta\theta$ of less
than a factor of 2.

\begin{figure}
\includegraphics*[width=8cm,height=5cm]{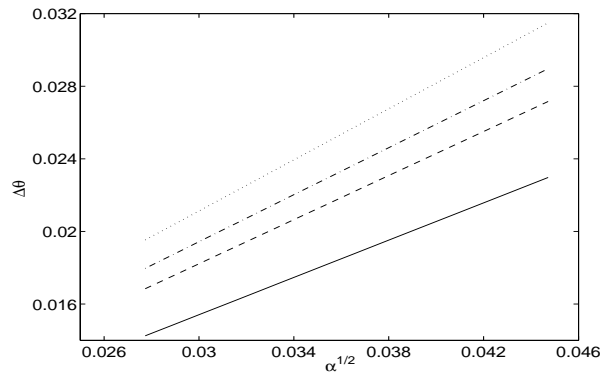}
\caption{Minimum value of $\Delta\theta$ as a function of
$\alpha^{1/2}$ for $\bar{n}_A=2.5$ and $\bar{n}_B=0$ (solid line), $\bar{n}_B=1$
(dashed line), $\bar{n}_B=1.25$ (dashed dot line), and $\bar{n}_B=1.5$ (dotted
line).}
\label{fig4}
\end{figure}

Fig. 5 provides a direct comparison of $\Delta\theta$ for the BEC
and Fermi beams when $N_B=0$. Remarkably, the phase resolution
limit achievable with the Fermi gas is noticeably better than what
can be achieved with even a noninteracting BEC. Moreover,
increasing the fermions density enhances the phase resolution. Of
course, this increase comes at the expense of a limited range of
$\theta$ over which the interference pattern is discernible. For
BECs, by contrast, $\Delta\theta$ is independent of $\theta$ for
$\beta=0$ and even for $\beta\neq 0$, $\Delta\theta$ is nearly
constant except in the vicinity of the resonances.

\begin{figure}
\includegraphics*[width=8cm,height=5cm]{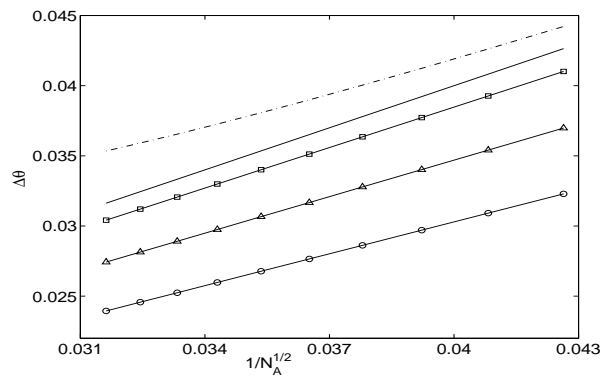}
\caption{Minimum value of $\Delta\theta$ for $N_B=0$ and bosons with
$\beta=0$ (solid line), bosons with $\beta=0.001$ (dash dot line), fermions
with $\bar{n}_A=1$ (squares), fermions with $\bar{n}_A=2$ (triangles), and
fermions with $\bar{n}_A=3$ (circles).}
\label{fig5}
\end{figure}

In conclusion, we have examined the phase resolution limit for an atomic
Mach-Zehnder using degenerate Bose and Fermi gases. Our results indicate that
one can attain better phase resolution using Fermi gases at the expense
of a limited range of path differences over which there is a discernible
interference pattern.

This work is supported in part by the US Office of Naval Research,
by the National Science Foundation, by the US Army Research
Office, by the National Aeronautics and Space Administration, and
by the Joint Services Optics Program.


\begin{references}
\bibitem{berman} Paul R. Berman, "Atom Interferometry" (Academic Press, San
Diego, CA 1997).
\bibitem{clauser} J. F. Clauser, Physica B {\bf 151}, 262 (1988).
\bibitem{scullydowling} M. O. Scully and J. P. Dowling, Phys. Rev. A {\bf 48},
3186 (1993).
\bibitem{rotation1} T. L. Gustavson {\em et al.}, Class.
Quant. Grav. {\bf 17}, 2385 (2000).
\bibitem{gradiometer} J. M. McGuirk {\em et al.}, Phys. Rev. A {\bf 65}, 033608 
(2002).
\bibitem{rotation2} A. Lenef {\em et al.}, Phys. Rev. Lett. {\bf
78}, 760 (1997).
\bibitem{rotation3} T. L. Gustavson {\em et al.}, Phys. Rev. Lett.
{\bf 78}, 2046 (1997).
\bibitem{dowling} J. P. Dowling, Phys. Rev. A {\bf 57}, 4736 (1998).
\bibitem{yurke} B. Yurke, Phys. Rev. Lett. {\bf 56}, 1515 (1986).
\bibitem{torii} Y. Torii {\em et al.}, Phys. Rev. Lett. {\bf 61}, 041602 (2000).
\bibitem{gupta} S. Gupta {\em et al.}, Phys. Rev. Lett. {\bf 89}, 140401 (2002).
\bibitem{yurke2} B. Yurke, S. L. McCall, and J. R. Klauder, Phys. Rev. A {\bf
33}, 4033 (1986).
\bibitem{sakurai} J. J. Sakurai, "Modern Quantum Mechanics" (Addison-Wesley,
Reading, MA 1994).
\end{references}
\end{document}